\documentclass[12pt,a4paper]{article}
\usepackage{bezier}
\usepackage{epsfig}
\usepackage{amssymb}
\usepackage{graphicx}
\usepackage{psfrag}

\newcommand{\nc}{\newcommand}
\nc{\be}{\begin{equation}}
\nc{\ee}{\end{equation}}
\nc{\bea}{\begin{eqnarray}}
\nc{\eea}{\end{eqnarray}}
\newcommand{\av}[1]{\langle #1\rangle}

\nc{\eqn}[1]{{(\ref{#1})}}
\nc{\cA}{{\cal A}}
\nc{\cB}{{\cal B}}
\nc{\cC}{{\cal C}}
\nc{\cD}{{\cal D}}
\nc{\cE}{{\cal E}}
\nc{\cF}{{\cal F}}
\nc{\cG}{{\cal G}}
\nc{\cH}{{\cal H}}
\nc{\cI}{{\cal I}}
\nc{\cJ}{{\cal J}}
\nc{\cK}{{\cal K}}
\nc{\cL}{{\cal L}}
\nc{\cM}{{\cal M}}
\nc{\cN}{{\cal N}}
\nc{\cO}{{\cal O}}
\nc{\cP}{{\cal P}}
\nc{\cQ}{{\cal Q}}
\nc{\cR}{{\cal R}}
\nc{\cS}{{\cal S}}
\nc{\cT}{{\cal T}}
\nc{\cU}{{\cal U}}
\nc{\cV}{{\cal V}}
\nc{\cW}{{\cal W}}
\nc{\cX}{{\cal X}}
\nc{\cY}{{\cal Y}}
\nc{\cZ}{{\cal Z}}
\nc{\tr}{{{\rm tr}\,}}

\nc{\bk}{{{\bf k}}}
\nc{\bx}{{{\bf x}}}

\nc{\simo}[1]{{\stackrel{#1}{\simeq}}}
\nc{\geqo}[1]{{\stackrel{#1}{\geq}}}
\nc{\geo}[1]{{\stackrel{#1}{>}}}
\nc{\guo}[1]{{\stackrel{#1}{\succ}}}

\nc{\rbo}{\raisebox}
\nc{\RR} {\rangle \! \rangle}
\nc{\LL} {\langle \! \langle}
\nc{\rmi}[1]{{\mbox{\small #1}}}
\nc{\eq}{eq.~}
\nc{\nr}[1]{(\ref{#1})}
\nc{\ul}{\bf }
\nc{\mc}{\multicolumn}
\nc{\todo}[1]{\par\noindent{\bf $\rightarrow$ #1}}

\nc{\cu}{{\cal u}}
\begin{document}
\begin{flushright}
SPhT-00/167 \\
BI-TP 2000/40\\
TPJU-15/2000 \\
HD-THEP-00-62
\end{flushright}
\centerline{\bf QCD with Adjoint Scalars in 2D:
Properties in the Colourless Scalar Sector}

\begin{center}  
P. Bialas$^{1,a}$, A. Morel$^{2,b}$, B.Petersson$^{3,c}$, K. Petrov$^{3,d}$
and T. Reisz$^{2,4,e}$
\end{center}

\centerline{$^{1}$Inst. of Comp. Science, Jagellonian University} 
\centerline{33-072 Krakow, Poland}
\vspace{0.3cm}
\centerline{$^2$Service de Physique Th\'eorique de Saclay, CE-Saclay,} 
\centerline{F-91191 Gif-sur-Yvette Cedex, France} 
\vspace{0.3cm}
\centerline{$^{3}$Fakult\"at f\"ur Physik, Universit\"at Bielefeld} 
\centerline{P.O.Box 100131, D-33501 Bielefeld, Germany}
\vspace{0.3cm}
\centerline{$^4$Institut f\"ur Theor.~Physik, Universit\"at Heidelberg}
\centerline{Philosophenweg 16, D-69120 Heidelberg, Germany}
\vspace{0.3cm}

\begin{abstract} \normalsize
We present a numerical study of an SU(3) gauged 2D model for adjoint scalar 
fields, defined by dimensional reduction of pure gauge QCD in (2+1)D at high 
temperature. We show that the correlations between Polyakov loops are
saturated by two colourless bound states, respectively even and odd
under the $Z_2$ symmetry related to time reversal in the original
theory. Their 
contributions (poles) in correlation functions of
local composite operators $A_n$ respectively of degree $n=2p$ and $2p+1$ in the scalar fields
($p=1,2$) fulfill factorization. The contributions of two particle states 
(cuts) are detected. Their size agrees with estimates based on a 
meanfield-like decomposition of the $p=2$ operators into polynomials in
$p=1$ operators. In contrast to the naive picture of Debye screening, no sizable signal in any $A_n$ correlation can be attributed
to $1/n$ times a Debye screening length associated with $n$
elementary fields. These results are quantitatively consistent with the 
picture of scalar ``matter'' fields confined within colourless boundstates 
whose residual ``strong'' interactions are very weak.
\end{abstract}

\vfill

\noindent
$^a$pbialas@agrest.if.uj.edu.pl \\
$^b$morel@spht.saclay.cea.fr \\
$^c$bengt@physik.uni-bielefeld.de \\
$^d$petrov@physik.uni-bielefeld.de \\
$^e$t.reisz@thphys.uni-heidelberg.de \\

\newpage

\section{\bf Introduction } 
Dimensional reduction is a powerful technique to study the infrared
region of field theories at high temperature\cite{ginsparg}--\cite{reisz}.
Combined with a non-perturbative lattice simulation of the
reduced model, it has been employed to investigate the properties
of gauge theories and QCD with dynamical quarks in the plasma phase
\cite{lacock}--\cite{kajantie1}. For a recent review
see \cite{philipsen}.

It is, however, still not clear, what are the limitations and the domain
of validity of this approach. Therefore, in
a recent work \cite {I}, we have studied  in detail the
reduction to two dimensions of pure gauge QCD in (2+1)
dimensions at high temperature. We refer the reader to this article for 
a more complete discussion of our
motivations, references to the related literature 
and details on the reduction procedure. The reduced model is a model for
scalars  belonging to the SU(3) algebra (formerly the electric
gluons in a static gauge). They interact with the 2D 
gauge fields, the static parts of the original 
3D spatial gauge fields,
and via an effective potential whose self-couplings are computed by 
perturbative integration over the non-static degrees of freedom. 
In \cite{I} we restricted the perturbative integration to the
one loop order. The main conclusion of our investigation
is that dimensional reduction works very well in this 
case. In particular it was shown that it works within a few percent for
the correlation function of Polyakov loops (as well as for spacelike
Wilson loops) down to $1.5 T_c$, where $T_c$ is the critical temperature
in the $(2+1)D$ theory. In fact for $T \geq 1.5 T_c$ numerical simulations
showed that the Polyakov loops
correlations measured in (2+1)D QCD and in the reduced model were
very close to one another down to
quite short distances, although our formalism is in principle an expansion
both in high temperature and in $p/T$, where $p$ is the relevant momentum
scale.
It was further shown, by simulations at different
values of the bare parameters, that our measurements are in the scaling
region, and thus our results are valid in the continuum
limit.

Less expected was the observation that these correlations
assume a shape typical of {\it single} particle propagation, 
as opposed to the standard picture of a screening mass associated with
{\it two} massive electric gluons. This feature, together with our previous  
motivations, invites us to pursue the numerical exploration of the reduced 
model per se, in particular the investigation of states connected with
the Polyakov loop correlations.

While in \cite{I} 
we only measured the correlations of Polyakov  
loops, here we analyze separately those of 
SU(3) invariant polynomials of degree $n$ in the elementary scalars
$A$, namely $A_n=\tr A^n$. The action is invariant under a global 
sign reversal of all the $A$'s. This $Z_2$-symmetry, which we denote by $R_{\tau}$,
following the authors of Ref. \cite{arnold}, corresponds to Euclidean
time reversal in the $(2+1)D$ theory.
In this latter article it was
suggested to investigate operators odd under this symmetry, to obtain
a possible gauge invariant definition of the Debye mass. 
To investigate both operators
which are even and odd under $R_{\tau}$
we consider separately correlations involving even and odd polynomials,
$n=2p$ and $2p+1$. The $R_{\tau}$ -symmetry may be spontaneously broken
in the $2D$ model. In fact there exists two phases, corresponding to
$R_{\tau}$ being conserved or broken. Only in the symmetric phase,
the model corresponds to the reduction of the high temperature
$(2+1)$ QCD phase
\cite{kark},\cite{thomas2},\cite{arnold},\cite{kajantie1}.
The  details of the actual phase diagram will not be studied 
in the present article. From our data we can conclude, however, that the 
values of the coupling constants in the reduced $2D$ theory are in
the unphysical broken phase. In the same way as has been done for the reduction
from $(3+1)D$ to $3D$, we solve this problem by working in the metastable part
of the symmetric phase.
Using zero field initial conditions on large enough 
lattices, we make sure that we stay in the phase of unbroken $R_{\tau}$,
where even and odd operators do not mix. 
Investigations of states in the full and reduced model
in the case of the $(3+1)\to 3$ reduction can be
found in \cite{lacock}--\cite{kajantie1} and \cite{kajantie3}--\cite{hart2},
although a detailed analysis of the nature of these states, as we
perform in this paper, have not yet been made.

The 2D action under consideration is recalled in Section 2, together
with its meaning in terms of the (2+1)D QCD model, from which it originates,
and the relevant operators and correlations are defined.
The simulations are performed in the temperature range $2T_c$ to $12T_c$,
where $T_c$ is the deconfining temperature of the latter model. 
Our results are presented and discussed in Sections 3 and 4. In Section 3,
we first show that the measured correlations fulfill the factorization
properties expected if the lowest states in the $n-$even and $n-$odd 
channels are two distinct one particle states, whose masses are then
extracted from the large distance decays. According to the criteria proposed
in Ref.\cite{arnold}, the state found in the odd channel is a candidate
to {\it define} a Debye screening mass, although not the only one.
In Section 4, we 
further analyze the composite operators $A_n,\,n=2p$ and $2p+1$ and their
correlations $A_{n,m}$, showing that all the properties observed for $p=2$
can be deduced with a good accuracy from their knowledge for $p=1$. This
follows from the assumption that, given the SU(3) and $R_{\tau }$ symmetry
constraints, the effective model for the elementary $A$'s after integration 
over the gauge fields is a free field model for the massive
composites $A_2$ and $A_3$. In particular, we give evidence that 
the (small) deviations from factorization observed at short distances are
mainly due to intermediate states containing two of the above particles.
The summary and the conclusions can be found  in a last Section 5.

\section {\bf The Reduced Action, Operators and Correlations} 

In this section, we write down the reduced 2D action 
derived in \cite{I}, and define our notations and
 the quantities of interest for the
present work.

The lattice is an $L_s\times L_s$ square; the spacing is $a$, set to one
unless specified otherwise. The weight in the partition function 
is written $\exp (-S)$, with $S$ a function of the SU(3) group
elements $U(x;i),\, i =1,2$ on the links and of the scalars $A(x)$ in the adjoint representation on the
sites:
\be \label{trace}
  A(x) = \sum_{\alpha =1}^{8} A^\alpha (x)\lambda^\alpha 
 , \quad \tr \lambda^\alpha  \lambda ^\beta  =  \frac {1}{2} 
 \delta_{\alpha \beta}.
\ee
Greek superscripts on $A$ will always denote colour indices, unlike
integers $n,m$ used in powers of the algebra element $A$. 
We write the 2D reduced action as follows:
\bea 
&&S=S_{U}+S_{U,A}+S_{A}, \label{s}\\
&&S_U = \beta_3 L_0 \sum_x 
   \biggl( 1 - \frac{1}{3} \Re\, \tr U(x;1) 
    U(x+a\widehat 1;2)
    U(x+a\widehat 2;1)^{-1} U(x;2)^{-1} 
    \biggr), \nonumber\\
&&S_{U,A} =\frac {\beta_3 L_0}{6}\sum_{x} \sum_{i=1}^{2}
    \tr \biggl( D_{i}(U) A (x) \biggr)^2, \label{sua}\\
&&D_{i}(U) A (x) = U(x;i) A (x+a 
     \widehat i) U(x;i)^{-1} - A(x), \nonumber \\
&&S_{A} = \sum_{x}
     k_2\,\tr A ^2(x)
      + k_4 \left( \tr A^2(x) \right)^2. \nonumber
\eea

In the above, $S_U$ is the pure gauge term, $S_{U,A}$ the gauge 
invariant kinetic term for the scalars and $S_A$ the scalar potential, whose 
self couplings $k _2$ and $k _4$ result from the one-loop
integration over the non-static components of the 3D gauge fields. All
terms have the global $R_{\tau}$-symmetry $A(x)\to -A(x)$, while the $Z_3$ symmetry of
the original $(2+1)D$ $SU(3)$ model is broken by the perturbative reduction
procedure.
 It was found
in \cite{I} that 
\bea  \label{k}
 k _2 &=& -\frac {3} {2\pi} \biggl (c_0\log L_0+c_1
 \biggr ); \,\quad c_0=1,\quad c_1=\frac {5}{2}\log 2-1,  \\
 k _4 &=& \frac {L_0^2} {64 \pi}. \nonumber
\eea 
The values of the
parameters $\beta _3,\,L_0$ follow from the original lattice 
regularization of 3D pure QCD at temperature $T$ and gauge coupling 
squared $g_3^2$ in the continuum. The latter has dimension one in energy 
and is used to set the scale:  
\bea \label{scaling}
\beta _3&=&\frac {6}{ag_3^2}, \\
L_0&=&\frac {1}{aT}. \nonumber
\eea
Accordingly, the continuum limit $a\to 0$ is obtained by letting
$\beta _3$ and $L_0$ go to infinity  with the dimensionless temperature
\be \label{tau}
\tau =\frac {T}{g_3^2}=\frac {\beta _3}{6L_0}
\ee
being kept fixed. 

The original three dimensional pure $SU(3)$ gauge theory has a global
$Z_3$ symmetry. The corresponding order parameter is the Polyakov loop.
It is a static operator. In the reduced theory it has the form
\be
L(x)=\frac {1}{3}\tr \exp\bigl [i\,L_0\,A(x)\bigr ]. \label{wline}
\ee
At sufficiently high temperature the symmetry is spontaneously broken,
signalling the deconfinement of static charges in the fundamental
representation. The reduced theory in the above form does not have
the $Z_3$ symmetry any more, because the perturbative reduction is 
made around one of the broken vacua, where $A(x)=0$. The phase transition
in the three dimensional theory appears at $\tau_c\simeq 0.61$\cite{lego}.
The reduced model should be valid in the deconfined phase, at sufficiently
high temperature and long distances.
In \cite{I} we employed it to investigate the correlations between
Polyakov loops in this phase, where they are related to screening. We performed a
detailed numerical analysis in the reduced model
and a comparison with the results in the full $(2+1)D$ theory.
Our simulations were performed for two
values of the parameter $L_0$, namely 4 and 8. It was shown that scaling was
very good, when comparing the data collected for fixed $\tau$ at these two values
of $L_0$. In this paper, we will show data collected for $L_0=4$. As this is
in the scaling domain, we can give the results in physical units. Distances
$R$ are given by
\be
RT=r/L_0,  \label{rphys}
\ee
where r is the distance in lattice units, and temperatures are measured
in units of the three dimensional critical temperature $T_c$. For $L_0$ 
fixed in the scaling region one may use
\be
T/T_c = \beta_3/\beta_{3c}. \label{tphys}
\ee

To discuss the effective Lagrangian of the model at fixed $\tau$ 
in the scaling limit, it is
convenient to normalize the scalar 
fields differently, defining $\phi (x)$ via  
\be
A(x)= \phi(x) {\sqrt \frac {6} {L_0\,\beta _3}}. \label
{phi} \\
\ee
The corresponding 
 Lagrangian $\cL _{eff}$ was derived in \cite{I} from the small $a$
  expansion of the effective action $S$. For clarity of the discussion
  we reproduce it here below. In $S$, we define $A_i$ by
\be
   U(x;i) = \exp[i\,a g_2A_i(x)], 
\ee
where $g_2^2=g_3^2T$ is the effective coupling of the 2D theory.
 Taking the limit $a\to 0$ (but for the UV logarithm $\log L_0\equiv -\log aT$ 
 in the $\phi ^2$ term), one obtains
 \bea
 \cL _{eff}&=&\frac {1}{4} \sum _{c=1}^8 F_{ij}^c\,F_{ij}^c + \tr
[D_i\phi]^2 +\frac {g_2^2}{32\pi}\biggl (\frac {g_2}{T}\biggr )^2\,
\tr \,\phi ^4 +\cL_{CT}, \\
 D_i \phi &=& \partial_i \phi + ig_2 [A_i,\phi],
\nonumber \\
 F_{ij} &=& \partial_i A_j - \partial_j A_i + i g_2 [A_i, A_j], \nonumber \\
 \cL _{CT} &=& -\frac {3g_2^2}{2\pi}
 \biggl [-\log (aT)+5/2\log 2-1\biggr ]\,\tr \,\phi ^2. \label{counter}
 \eea

This is a $2D$, $SU(3)$ gauge invariant Lagrangian for an adjoint scalar
$\phi$, but it is far from being the most general one. The gauge coupling
$g_2$, with its canonical dimension one in energy, sets
the scale. The non kinetic quadratic term is
the counterterm $\cL_{CT}$, suited to a lattice UV regularization with spacing
$a$. The appearance of this term in the context of dimensional reduction 
in e.g. the lattice regularization framework was
first discussed in Refs.\cite{nadkarni}\cite{reisz}. It is well known that such
logarithmic terms also appear in general, when one wants to define the continuum
limit of a $2D$ lattice model, see e.g.\cite{parisi}\cite{gleiser}.  
The reflection symmetry $\phi \to -\phi$ 
is related to the euclidean time reflection in the original $(2+1)D$ theory,
noted $R_{\tau}$ and discussed by Arnold and Yaffe \cite{arnold}. 
In the two dimensional model of Eq. (\ref{s}), it may,
however, be spontaneously broken  in some subspace of the unrestricted parameter
space  {$\beta _3L_0,\, k_2,\,k_4$}. As was discussed in Refs. \cite{kark} \cite{thomas2}
\cite{kajantie1}, and as follows from the invariance of the original 
three dimensional theory under euclidean time reflection,
the physical phase is the $R_{\tau}$ symmetric phase.
The phase diagram in the $3D$ adjoint Higgs model, related to
$4\to 3$ QCD reduction
has been studied in Ref. \cite {kajantie2}.
For the case of SU(2) in (3+1)D, $R_{\tau}$ is
the center of the gauge group so that $R_{\tau}$ breaking is also gauge symmetry
breaking, a subject previously discussed in Refs. \cite {kark}, \cite
{thomas2} and \cite{kajantie1}.

We have made a numerical investigation of the relative positions of the
phase transition and of the reduction point in the reduced model, 
for $T/T_c=1.97$,
and for two values $L_0=4$ and 8. We find that the reduction point is near to the phase transition, but in the broken phase. The phase transition is 
first order and strong enough, so that we can study the reduction in the metastable
symmetric phase, by using appropriate starting values for the fields, and employing
a large enough $(32\times 32)$ lattice.

We now turn to the definitions of the quantities relevant for
the present investigation of the 2D model. 
The Polyakov loop correlation $\cP(x)$ is defined by
\be
\cP(x)=\av{L(0)\, L^\dagger(x)}-|\av{L}|^2,  \label{plc}
\ee
where $L(x)$ is the Polyakov loop operator of Eq. (\ref{wline}).
We now expand $L(x)$ in powers of $A(x)$, and we will study the operators
$A^n(x)$ and connected correlations of their traces, defining
\bea
A_n(x)&\equiv&\tr {A^n(x)}, \label{an} \\
A_{n,m}(x)&\equiv&\av {A_n(x)\,A_m(0)}\,-\,\av{A_n}\,\av {A_m}. \label{anm}
\eea
When not ambiguous, the notation $A_n$ may also represent $\av {A_n(x)}$.
Any operator $A_n(x)$ is gauge invariant, even or odd under the
$R_{\tau}$-symmetry of $S$ for $n$ even or odd respectively.
Because the reduced model was derived from a {\it small} $A$-fields expansion, 
its properties are significant for the (2+1)D model in the unbroken
$R_{\tau}$ phase only, where $ A_{2p+1}=0$ and $A_{2p}$ is small.  

In this article we will concentrate on the study of correlations of these
operators, which are directly related to the static operator $L(x)$ and therefore
relevant for the high temperaure properties of the original $(2+1)D$ $ SU(3)$ theory.
As can be easily seen from the effective Lagrangian above, the $2D$ model has
further symmetries beside $R_\tau$, namely reflections of the 1- and
2-axis, which can be used
to classify further operators, whose correlations may be studied. 
(In $2D$ there is no spin quantum number.)
A corresponding
analysis has been made in \cite{kajantie2} \cite{hart2} in the case of the 
three dimensional
adjoint Higgs model. Although
a full numerical analysis of the two dimensional adjoint Higgs model certainly has
an interest {\it per se}, the operators other than those which we defined above
are not directly related to a {\it static} $(2+1)D$ operator. One would need a 
further 
study to acertain to what extent their correlations are related to the high 
temperature physics of the original theory. We therefore do not discuss them 
in the context of this paper.
\vskip 0.3cm

We have performed a numerical simulation of the model defined above,
with a flat measure for the $A$'s and the standard De Haar measure
for the gauge fields. The algorithm and error estimate techniques used
are the same as in \cite{I} and not reproduced here. The lattice
size is $L_S=32$, and $L_0=4$, throughout the present work. The $\beta
_3$ values are 29, 42, 84 and 173. This corresponds to
values of $T/T_c$ equal approximately to 1.97, 2.85, 5.70 and 11.73 
respectively. We were able
to extract information from operators and correlations corresponding to
$n=2$ to 5. The cases $n=2$ and 3 for the $3D$ reduced model were 
investigated in \cite{kajantie2}. 

\section {\bf  The Lowest States of the Scalar Spectrum }
In this section we present our results for the $A_{n,m}$ correlations
measured, and describe them for each temperature in terms of two states
$S$ and $P$, respectively even and odd under $R_\tau$ and appearing for
for $n,m$ both even and both odd. Their physical masses will 
be denoted $M_S$ and
$M_P$. For simplicity we often use in this and the next section
the bare parameters $r$ and $\beta_3$, related to $R$ and $T$ by Eqs. (\ref{rphys}) and (\ref{tphys}). \\

In the simulations reported here, all runs were initialized with
zero $A$-fields values and, as already stated, the system stayed in the 
metastable $R_{\tau}$ unbroken phase, as desired, with $\av {A_{2p+1}(x)}$ being always compatible with zero. In this respect, the situation is thus similar
to that encountered in (\cite{kark}-\cite{kajantie2}).

For a first look at the data obtained in
even and odd channels, we show in Fig.~1 the on-axis correlations
$A_{n,m}(r),\,\,n\leq m\in [2,5]$
at $T/T_c=1.97\,(\beta _3=29)$. They are plotted against $RT$, that is the 
physical distance in units of the inverse temperature.
In the even cases, the three correlations
all have the same shape, and they decrease by about one
order of magnitude each time two more powers of $A$ are involved. The
same is true for the odd cases, with a common decay of 
the correlations steeper than in the former case (smaller correlation length
in lattice units). The overall situation is similar for $\beta _3$ higher. 
For $n,m$ larger than 5, as well as for $\beta _3$ very large, the
signal/noise ratio becomes very small. This can be understood at the
qualitative level by noting that the rescaling Eq.(\ref {phi}) of
the $A$-fields  normalizes the kinetic term for the $\phi$-fields
to the standard, parameter independent form $1/2\,\tr {(D_i\phi )^2}$.
Hence if the field renormalization by the interactions is weak, the  $\phi$
correlations should depend only  weakly on $\beta _3$, which means 
that $A_{n,m}$ scales like $\beta _3^{-(n+m)/2}$.
This will be illustrated more 
quantitatively in section 4.
Due to this scale factor, the $A$-fields remain ``small'' in practice
down to quite low values of $\beta _3$, which a posteriori explains why
the perturbative reduction may still work at a temperature as low as 1.5
$T_c$. In fact, we checked that the Polyakov loop correlations are
actually fully reconstructed within errors by keeping $\{n,m\}$ up to
$\{5,5\}$
only in their expansion in $A_{n,m}$'s obtained from the small $A$
expansion of (\ref{plc}).

Now we want to analyze these $A_{n,m}$ data quantitatively in terms of the
lowest states of the spectrum. Let $m_i=a\,M_i$ be the lowest
mass with quantum number $i=S,P$, in lattice units. For particle $i$ we 
introduce a lattice propagator $\widetilde \Delta _{Latt}(m_i,p)$ 
in momentum space: 

\bea  \label {proplatt}
\widetilde \Delta ^{-1}_{Latt}(m_i,p)=
\widehat {p}^2+4\sinh (m _i^2/4),\\
\widehat {p}^2=4\sin ^2(p_1/2)+4\sin ^2(p_2/2). \nonumber
\eea
The corresponding contribution to $A_{n,m}(r)$ then reads
\be
A_{n,m}(m_i,r)=g^{i}_{n,m}\frac {1}{L_s^2} \sum _{p_1,p_2} \cos(p_1r) 
\widetilde \Delta _{Latt}(m_i,p), \label{onaxis} \\
\ee
where $g^i_{n,m}$ measures the residue of $A_{n,m}$ at the pole of
(\ref{proplatt}), which on large enough lattices sits at 
$p^2\sim \widehat p^2\sim -m_i^2$.  With our definitions,
$g^i_{n,m}$ is non zero only for $i=S$ if $n$ and $m$ are even, and for
$i=P$ if $n$ and $m$ are odd. {\it Prior to any fit of the masses to the
data}, we notice that a first consequence of our expectations on the
lowest part of the spectrum comes from residue factorization $g^i_{n,m}=\gamma
^i_n \gamma ^i_m$, a property which we can probe directly on the
correlations since, as $r$ becomes large, it implies  
\be \label{factor}
X_n\equiv \frac {A_{n,n}(r)\,A_{n+2,n+2}(r)}{A_{n,n+2}^2(r)}\to 1.
\ee
That it is so is demonstrated on Figs. (2-5) showing $X_2$ and $X_3$ 
(symbols $\diamond$) for $T/T_c=1.97$ and 5.7 ($\beta _3=29$ and 84). The agreement is very good in
all cases, although the quality of the data is poorer for 
$X_3$ due to the correlations involving $A_5$ getting very small. 
Similar results are obtained for other values of $T/T_c$. 
We thus conclude at this point that single particle propagation accounts
very well for the largest correlation length occuring in each of the two
channels.  Most of the observed deviations of $X_n$ from one will be 
interpreted in the next section in terms 
of two particle state contributions (symbols $\circ$ in the same figures).

We now proceed to assign values to the two lowest masses $M_S$ and $M_P$
expected from the above findings. This we do by various ways, in order
to further enforce the statement that the correlations do have the 
characteristics associated with the pole structure of Eq. (\ref{proplatt}).
Down to $r\sim 1$, an excellent approximation to the on-axis correlation 
(\ref{onaxis}) is given by
\be \label{pole}
A_{n,m}(m_i,r)\simeq
c\,\Biggl (\frac{1}{[m_ir]^{1/2}}e^{- m_i\,r}+
\frac{1}{[m_i(L_s-r)]^{1/2}}e^{- m_i (L_s-r)}\Biggr ),
\ee
where $c$ is constant in $r$. We performed fits of this formula to 
all our $A_{n,m}(r)$ data taken at $r>r_{min}$. These fits are stable 
with respect to $r_{min}$ provided  it is larger than about 4, and the
values found for $m_i$ in different correlations are always consistent 
with each other. The smallest errors were obtained by using 
fits to $A_{2,2}$ and $A_{3,3}$. 

Effective masses can also be obtained without any fitting by
using $0-$momentum correlations, defined for a generic
$x-$space correlation $C(x_1,x_2)$ by
\be
C^0(r)=\frac {1}{L_s}\sum _{x_2}\,C(r,x_2). \label{prj}
\ee
If the lowest mass in $C$ is $m$, the ansatz (\ref{proplatt}) gives
\be \nonumber
C^0(r)\propto \cosh \biggl (m(L_s/2-r)\biggr ),\nonumber
\ee
and $m$ can be extracted at any $r$ by inverting this relation:  
\bea \label{meff}
m^{eff}(r)=\log \biggl(Y(r)+\sqrt {Y^2(r)-1}\biggr ), \\
Y(r)=\frac {C^0(r+1)+C^0(r-1)}{2C^0(r)}. \nonumber
\eea
As an overall consistency check, we have extracted an effective mass 
$m^{eff}(r)$ from $0-$momentum Polyakov loops correlations (\ref{plc}), 
and compared it to the $m_S$ values obtained by our fits to $A_{2,2}$. 
We find that $m^{eff}(r)$ is indeed nearly constant, in fact slowly 
decreasing towards a value compatible with $m_S$, due to
smaller and steeper contributions to (\ref{plc}) of the heavier particle $P$.
 
A contrario, we invalidate the interpretation of the largest
correlation length in $A_{n,n}$ as being $n$ times shorter than the 
``Debye screening length'', the inverse of a mass 
$m_E$ associated with ``electric'' gluons of the initial (2+1)D model  
(the scalars of the reduced model). This scenario was advocated by D'Hoker 
in his perturbative study of $QCD_3$ at high temperature \cite{dhoker}. If such 
was the case, the on-axis correlations $A_{n,n}$ should rather look like
\be \label{cut}
A_{n,n}(n\,m_E,r)\propto \Biggl (\frac{1}{[m_Er]^{1/2}}e^{- m_E\,r}+
\frac{1}{[m_E(L_s-r)]^{1/2}}e^{- m_E (L_s-r)}\Biggr )^n,
\ee
which differs in shape from (\ref{pole}), as was illustrated in \cite{I}
for Polyakov loop correlations. We nevertheless tried fits with (\ref{cut}),
but got definitely worse agreement, in the range of temperatures,
which we have investigated, i.e up to $12T_c$. Hence the above picture is
ruled out by the data in this temperature range, and if a mass can be defined for the electric
gluon in high temperature $QCD_3$ it is most probably larger than both $m_S$/2 
and $m_P$/3. In a "constituent gluon" picture, as advocated in Ref. \cite{buchmuller},
one would have bound states instead of a cut. One would, however, expect
$m_P/m_S \approx 3/2$.

Our final results for the $S$ and $P$ masses in units
of the scale $\sqrt{g_3^2 T}$
are collected in Table \ref{tbl:msmp} for the values of
$T/T_c$ investigated.
They are taken from fits to $A_{2,2}$ and $A_{3,3}$
respectively. The values for $M_S$ agree with those obtained
in \cite {I} from the Polyakov loop correlations. As can be seen from the
tables, the ratios $M_P/M_S$ vary with $T/T_c$. There is, however, no clear 
tendancy in the region we have investigated, the ratios being 1.8, 2.0, 1.7,
1.6 in order of increasing temperature. We can, of course, not exclude that
the ratio goes to 1.5 at still higher temperatures.
\begin{table}
\begin{center}  
\begin{tabular}{||r||l|l||l|l||}
\hline\hline
$T/T_c$ & $M_S/\sqrt{g_3^2T}$ & $M_P/\sqrt{g_3^2T}$ \\\hline\hline
1.97 & 1.39(3) & 2.51(13) \\\hline
2.85 & 1.48(5) & 3.02(16) \\\hline
5.70 & 1.89(5) & 3.22(15) \\\hline
11.73& 2.12(8) & 3.44(11) \\\hline\hline
\end{tabular}

\end{center}
\caption{\label{tbl:msmp} Masses in units 
of $\sqrt{g_3^2 T}$ for the $S$ and $P$
states, as measured from fits to $A_{2,2}$ and $A_{3,3}$ respectively,
for different values of $T/T_c$.}
\end{table}
\section {\bf  Weak ``Strong'' Interactions Between Colourless States}

Here we will show that even at quite short distances ($r$ small compared to $m_S^{-1}$)
all the condensates $A_n\equiv \av {A_n(x)}$ and correlations $A_{n,m}(r)$ 
can be reconstructed to a good accuracy from the data for $A_2$,
$A_{2,2}$ and $A_{3,3}$. The assumption is that the elementary fields 
$A^\alpha (x)$ (Greek superscripts are colour indices) interact only through 
$S$ and $P$ exchanges between the non-interacting composite $A_2(x)$ and 
$A_3(x)$, the scale of the fields being fixed by the size of 
$A_2$, while $A_3$=0. The precise way how this idea is implemented and
the corresponding technicalities are detailed in the appendix.

Here we limit ourselves to the simplest applications and give the results,
starting with the local condensates. 

\subsection {Weak Residual Interactions: The $A$-fields condensates}

Since SU(3) has rank 2, any $A_n(x)$ can be reduced to a polynomial in
$A_2(x)$ and $A_3(x)$. An elegant method \cite{bauer} and explicit formulae are given in the Appendix. For
$n$ odd $A_n$ is zero by $R_{\tau}$ 
symmetry. For $n$ even, we apply Wick contraction to all pairs of $A^\alpha$
elementary fields,  followed by the meanfield-like substitution
\be \label{subst}
A^\alpha (x)\,A^\beta (x) \,\, \to\,\,\frac {1}{4}\delta_{\alpha ,\beta}\,
A_2(x).
\ee
As an illustration, consider $A_4$. With the definitions of Section 2,
we have 
\be \label{suma4}
2A_4(x)=A_2^2(x)=\frac {1}{2^2}\,\sum _{\alpha ,\beta=1}^8
A^\alpha (x)A^\alpha (x) A^\beta (x) A^\beta (x).
\ee
There we  apply (\ref{subst}) and then replace $A_2(x)$ by its average
$A_2$. The $A^\alpha A^\alpha $ and 
$A^\beta A^\beta$ contractions give $(8\times A_2/4)^2$, and the 
additional contributions from $\alpha =\beta$ give  $2\times 8(A_2/4)^2$.
Noting that $A_{2,2}(0)=\av {A^2_2(x)}-A_2^2\,$ (see (\ref{anm})), 
the net result can be put into the two equivalent forms 
\bea
2A_4&=&\frac {5}{4}\,A_2^2, \label{a4} \\
A_{2,2}(0)&=&\frac {1}{4}\,A_2^2. \label{a22}
\eea
This prediction is remarkably well verified in all cases. 
At $\beta _3 =29$, the left and right hand sides of (\ref{a22}) are 
respectively $4.86(1)\,10^{-3}$ and $4.825(10)\,10^{-3}$. They are 
$6.093(13)\,10^{-4}$ and $6.069(1)\,10^{-4}$ at $\beta _3=84$.
Similar manipulations lead to
\be \label{a33}
A_{3,3}(0)=\frac {5}{64} A_2^3.
\ee
In this case the left and right hand sides are measured to be 
$2.105(7)\,10^{-4}$ and
$2.095(6)\,10^{-4}$ for $\beta _3=29$, $9.365(30)\,10^{-6}$ and
$9.344(3)\,10^{-6}$ for $\beta _3=84$. Hence the effects of 
residual interactions via non quadratic effective couplings in $A_2(x)$ 
and $A_3(x)$ are less than the percent in the correlations at zero
distance.

Before going to the correlations at non zero distance, let us 
discuss their normalization, as measured by the values of $A_{2,2}(0)$
and
$A_{3,3}(0)$ described just above. At the beginning of section 3, we
argued that the behaviour in $\beta _3,\,n,\,m$ observed for $A_{n,m}$
could
follow from the absence of a large renormalization, by the
interactions, of the $\phi$-fields defined by
Eq. (\ref{phi}). Here we note that in the confined phase the effective
degrees of freedom  are the ${\it massive}$ composites
$\phi _i=\tr {\phi ^i},\,i=2,\,3$, so that in the limit where they are
considered as free fields, one may write (see (\ref{proplatt}))
\bea  \label{residue}
\av {\phi _i(0)\,\phi _i(0)}&\simeq& R_i\int d^2p\,\frac {1}
{\widehat {p}^2+4\sinh (m _i^2/4)}, \\
\widehat {p}^2&=&4\sin ^2(p_1/2)+4\sin ^2(p_2/2), \nonumber
\eea
the residue $R_i$ being one if neither $\phi$ nor the composites get
renormalized. We computed $R_i$ as the
ratio of the l.h.s. of (\ref{residue}), directly measured, to the
integral 
in the r.h.s, evaluated numerically on the lattice for the mass values
fitted to the correlation data. The result is shown in Fig.6: in the
whole temperature range, both residues in the even and odd channels
remain uniformly very close to one.

\subsection{Weak Residual Interactions: Properties of the Correlations}

As we have seen in section 2 (see Fig. 1), the different $A_{n,m}(r)$'s
corresponding to the same channel have very similar shapes. Their 
analysis in terms of one particle exchange was successful, confirmed 
by the agreement with residue factorization. Nevertheless although the quantity
$X_n$ of Eq. (\ref{factor}) does go to one at large distances, it is significantly 
different from one at medium and short distances (see Figs. 2-5).
We will now show that two particle exchange is responsible 
for most of this lack of factorization.

The simplest consequence of our assumptions for correlations at finite $r$ is, 
using Eq.~(\ref{a4}),
\be  \label{wicka24}
A_{2,4}(r)=\frac {5}{4}\,A_2\,A_{2,2}(r), 
\ee
which is very well verified at any distance as shown in Fig.7 for $T/T_c=1.97$
($\beta _3~=~29$). 
A similar agreement is found for the relation
\be
A_{3,5}(r)=\frac {35}{24}\,A_2\,A_{3,3}(r),
\ee
derived in the appendix.
A new situation arises when we consider $A_{4,4,}$ or $A_{5,5}$ where
both the initial and final states may couple to a two particle state, 
$(S\,S)$ or $(S\,P)$ respectively. Then the intermediate state in a connected
correlation between $0$ and $r$ may consist of either one or two particles.
For example, to compute $A_{4,4,}(r)$, we apply the substitution rule 
(\ref{subst}) to the sum (\ref{suma4}), and then average using the
definitions of $A_{2,2}$ and $A_2$. One finds
\be
4A_{4,4}(r)=\biggl (\frac {5}{4}\biggr )^2\,
\biggl [4A_2^2\,A_{2,2}(r)\,+2\,A^2_{2,2}(r)\biggr ].
\ee
A similar treatment given in the appendix leads to the prediction
\be
A_{5,5}(r)=\biggl (\frac {35}{24}\biggr )^2\,
\biggl [A_2^2\,A_{3,3}(r)\,+\,A_{2,2}(r)\,A_{3,3}(r)\biggr ].
\ee
In the two expressions above, the second contribution, a product of two
propagators in space, is that of a two-particle intermediate state, and 
it provides a correction to exact factorization. From the definitions 
Eq.(\ref{factor}) of $X_2$ and $X_3$, one actually gets the following
estimates:
\bea
X_2(r)&\simeq &\widetilde X_2(r)\equiv 1+\frac {A_{2,2}(r)}{2A_2^2}, 
\label{correct2}\\
X_3(r)&\simeq &\widetilde X_3(r)\equiv 1+\frac {A_{2,2}(r)}{A_2^2}. 
\label{correct3}
\eea
The estimates $\widetilde X_2(r),\,\widetilde X_3(r)$ are displayed
in Figs. 2,3 (resp. 4,5), for comparison with the measured values 
$X_2(r),\,X_3(r)$ at $T/T_c=1.97$ (resp. 5.7), i.e. 
$\beta _3=29$ (resp. 84). We see that
the corrections to factorization implied by two-particle 
propagation provide  a reasonable explanation of the behaviour
observed for the $X$'s at intermediate and short distances. This is
especially true in the case of $X_2$, showing that there is very little
room for contributions from direct non-quadratic couplings in $A_2(x)$
in the full effective action (that resulting from integration over the gauge 
fields). This justifies our statement that the residual interactions between 
the colourless boundstates of the adjoint scalars are very weak.

\section {\bf  Conclusions} 

In this paper, we have studied properties of the two dimensional model
derived in \cite{I} by dimensional reduction of 3D QCD at high temperature.
In this model, scalar fields $A$ in the adjoint representation of SU(3)
interacts via SU(3) gauge fields $U$, in addition to a self-interaction
generated  by integration over the non-static 3D gauge degrees of freedom.
 Such properties are interesting since it was shown in \cite{I}
that dimensional reduction works remarkably well in this case. Also,
they offer an opportunity to explore non-perturbative features in a
low dimensional situation where the IR  singularities are particularly 
severe. 

By means of numerical simulations, we explored that part of phase space 
where the $R_{\tau}$-symmetry $A\to -A$ is unbroken, in accordance with the small 
$A$ expansion used to derive the model, known to be valid quite soon
above the transition temperature of pure 3D QCD. We identified two boundstates 
$S$ and $P$, respectively even and odd under $R_{\tau}$, and thus
coupled to monomials respectively of degree $2n$ and $2n+1$ in the $A$'s. 
The $S$ signal coincides with that previously obtained from Polyakov loops
correlations \cite{I}, where however the $P$-state contributions could not 
be disentangled.

These results came out from the measurement of three even-even
and three odd-odd distinct correlations, as functions of the on-axis
lattice distance $r$. Great care was taken in the analysis of their shape in
$r$, with the result that in all cases, the signal found was
that expected from the occurrence of genuine poles in momentum space.
A contrario, this demonstrates that the picture where the decay with
$r$ of such correlations reflects the propagation in 3D of  
$p=2n$ or $2n+1$ ``electric gluons'', i.e. a correlation 
length equal to $1/p$ times the ``Debye screening length'', is
inadequate in the case under study.

By comparing the size of the three different correlations measured 
for each of the $S$ and $P$ sectors, we were able to 
show that residue factorization holds, as expected on general grounds
when one particle propagates between different states. The agreement
with factorization was expectedly found to be particularly good at large 
distances, but we could even show that deviations  
at shorter distances are to a large extent compatible with propagation
of two particles,  namely two $S$ or  $S$ and $P$ respectively in the
$S$ or $P$ channel. The overall picture thus is that the scalar sector of 
the reduced model at large distances, thought to accurately describe static 
properties of 3D QCD at high temperature, consists of two
weakly interacting colourless particles, respectively even and odd under the $R_{\tau}$ symmetry of the model.

There are several problems, which this study invites to investigate
further. Of course, similarly detailed analysis for full QCD in 
(3+1)D would be interesting. 
Furthermore, the construction of a 
reduced model where the $Z_3$-symmetry of the 
pure gauge theory is not spoiled by the
reduction process is highly desirable \cite{pisarski}, with the hope that 
it exhibits a transition to a symmetric $Z_3$ phase analogous to the low 
temperature QCD phase. 
\section {\bf  Acknowledgments} 
We thank the DFG for support under the contract Ka 1198/4-1. 
K.P. was also supported by DAAD and  P.B  partially 
by KBN grant P03B01917.  

\appendix
\vskip 2.5cm
\begin{center}
\large
{\bf Appendix: Mean Field Technique for Composite Fields Correlators}
\end{center}
\vskip 1.5cm
\normalsize
\section {Formulae for Traces and Determinants }

Let $\phi$ be a complex $N\times N$ matrix. Here  we give
an elegant trick \cite{bauer} to compute the traces 
\be
\phi_n\equiv \tr \phi^n
\ee 
for $n> N$, given $\phi_p$ for $p\leq N$.

Consider the determinant $P_N(t)\equiv Det(1-t\phi)$, where $t$ is a
complex variable. It is a polynomial of degree $N$ in $t$ and its term
of degree $N$ is $(-1)^NDet(\phi)$, and we have 
\be
   \log \biggl (P_N(t)\biggr ) \; = \;
    \tr \log \left( 1 - t \phi \right).  
\ee
Both sides of this identity can be expanded in $t$ in some finite
neighbourhood of zero. The method consists in identifying the coefficients of 
the two series. The $N$ first orders determine the
coefficients of $P_N(t)$ from the $\phi_n$'s $,n\leq N$.
Then, the higher orders directly express any $\phi_n$, $n> N$
as a function of the $\phi_p$'s, $p\leq N$. Note that instead
of computing $Det(\phi)$ from the order $N$ coefficient
of $P_N$, one can alternatively compute $\phi_N$ given
$Det(\phi)$, which is convenient for SU(N) group matrices.

\vskip 0.5cm  
If applied to $\phi=A$, an element of the SU(3) algebra,
(in which case $A_1=0$), this technique gives $A_n,\, n>3$ in
terms of $A_2$ and $A_3$ taken as independent variables. The first non
trivial identities are

\bea
   A_4 & = & \frac{1}{2} (A_2)^2,
   \nonumber \\
   A_5 & = & \frac{5}{6} A_2 A_3,\label{a5} \\
   A_6 & = & \frac{1}{4} (A_2)^3 
              + \frac{1}{3} (A_3)^2, 
   \nonumber \\
   A_7 & = & \frac{7}{12} A_3 (A_2)^2, 
   \nonumber \\
   A_8 & = & \frac{1}{8} (A_2)^4 
              + \frac{4}{9} (A_3)^2 A_2, \nonumber
\eea
and the determinant is
\be
   Det A \; = \; \frac{2A_3}{3!} .
\ee
\vskip 0.5cm  
In what follows, we will have to manipulate monomials of the 
elementary scalar fields $A^\alpha (x)$, defined for SU(N) through 
\be
  A(x) = \sum_{\alpha =1}^{N^2-1} A^\alpha (x)\lambda^\alpha 
\ee
where the traceless basis $\lambda^\alpha$ is subject to the normalization
\be
\tr \lambda^\alpha  \lambda ^\beta  =  \frac {1}{2}
 \delta_{\alpha \beta}. 
\ee
On this basis the anti-commutators read
\be \label{anticom}
   \{ \lambda^\alpha, \lambda^\beta \} \; = \;
     c_{\alpha \beta} \; 1_N \; + \;
      \sum_{\gamma=1}^{N^2-1} d_{\alpha \beta \gamma} \; \lambda^\gamma ,
\ee 
with real and totally symmetric tensors $c$ and $d$. With these
normalizations and notations, we have 
\bea 
\tr A^2&=&\sum ^{N^2-1}_{\alpha \beta =1}\tr [\lambda^\alpha \lambda^\beta ]
\,A^\alpha A^\beta 
=\frac{1}{2}\,\sum ^{N^2-1}_{\alpha \beta =1}
\,\delta_{\alpha
\beta}\,A^\alpha A^\beta, \label{tr2} \\
\tr A^3&=&\sum ^{N^2-1}_{\alpha \beta \gamma =1}\,
\tr [\lambda^\alpha \lambda^\beta \lambda^\gamma ]\,
A^\alpha A^\beta A^\gamma 
=\frac{1}{4}\sum ^{N^2-1}_{\alpha \beta \gamma =1}\,
d_{\alpha \beta \gamma}\,
A^\alpha A^\beta A^\gamma . \label{tr3}
\eea

The projection property 
\be
   \sum_{\alpha=1}^{N^2-1} \lambda^\alpha_{ab} \lambda^\alpha_{dc}
   \; = \;
   \frac{1}{2} 
   \left( \delta_{ac} \delta_{bd}
      - \frac{1}{N} \delta_{ab} \delta_{cd}
   \right) 
\ee

can be used to derive that for any pair $X,Y$ of complex $N\times N$ matrices 
the following identities hold:
\bea \label{prod_to_trace}
    \sum_\alpha \tr \lambda ^{\alpha} X \; \tr \lambda ^{\alpha} Y \; &=& \;
   \frac{1}{2}\, 
   \left( \tr XY \; - \; \frac{1}{N} \; \tr X \; \tr Y
   \right),\\
   \label{trace_to_prod}
    \sum_\alpha \tr X \lambda ^{\alpha} Y \lambda ^{\alpha} \; &=& \;
   \frac{1}{2}\, 
   \left( \tr X \; \tr Y \; - \; \frac{1}{N} \; \tr XY \right). 
\eea

In what follows we specialize to $N=3$.

\section{Correlators of Composite Operators}

By gauge invariance, $A_2(x)$ and $A_3(x)$ 
can be chosen as the two effective degrees of freedom. By $R_{\tau}$-symmetry, 
the even and odd sectors under $A\to -A$ decouple. Here we derive consequences 
of the assumption that their dynamics is determined at leading order by their 
given vacuum expectation values $A_2$ and 0 respectively and their connected
two-body correlations $A_{2,2}(x)$ and $A_{3,3}(x)$. 

We use a Wick like treatment to express the higher order connected
correlation functions and averages through the quantities mentioned
above.
If $n=2p$, any $A^\alpha (x)$ is 
assigned to belong to a pair $A_2(x)$, then considered as a free field denoted
$S(x)$. So each monomial is replaced by a sum over all such pairings,
and each of the $p$ pairs is subject to the substitution $W_2$,
\be 
W_2:\quad \qquad \; A^{\alpha } A^{\beta } \to
   \delta_{\alpha  \beta } \; \frac{1}{4}\,S(x), 
\ee
leading to a monomial of degree $p$ in $S(x)$. If $n=2p+1$, one first
performs the $p-1$ possible substitutions $W_2$ 
(the result after $p$ substitutions would transform as an octet under
$SU(3)$ and thus vanishes), which yield a monomial necessarily proportional 
to $S^{p-1}(x)\,\tr A^3(x)$. There we apply the substitution $W_3$,

\be 
W_3:\qquad   \tr {A^3(x)} \to \,P(x),
\ee
where $P(x)$ is also considered as a free field. 

Once the local operators have been expressed in terms of $S(x)$ and $P(x)$, 
any average is obtained by using
\bea 
\av{S(x)} &=& A_2, \label{avs}\\
\av {S(x)\,S(0)} &=& A_{2,2}(x)+ A_2^2, \label{wick2} \\
\av{P(x)\,S(0)} &=& 0, \label{avps} \\
\av{P(x)} &=& 0, \label{avp} \\
\av{P(x)\,P(0)} &=& A_{3,3}(x) \label{wick3}. 
\eea
These rules generalize the way how in Section 4 we computed averages involving
$A_4(x)$. Let us now detail calculations involving $A_5(x)$. 

From Eqs.(\ref{a5}, \ref{tr2}, \ref{tr3}), we have 
\be
\frac{6}{5}\,A_5(x)=\sum _{\alpha \,\beta } 
\, A^\alpha A^\beta\,\, \tr \lambda^\alpha \lambda^\beta 
\,\sum _{\gamma ,\delta ,\epsilon}\,
\, A^\gamma A^\delta A^\epsilon \,
\tr \lambda^\gamma \lambda^\delta \lambda^\epsilon.
\ee
We apply rule $W_2$ to the right hand side. The contraction 
of $\alpha$ with $\beta$ produces $S(x)\,P(x)$ once. Using Eqs.
(\ref{prod_to_trace}, \ref{trace_to_prod}, \ref{tr2}, \ref{tr3}), one 
finds that the 6 $W_2-$contractions of either $\alpha$ or $\beta$ 
with either one of the three other indices contribute each the same amount
\be
\frac{1}{4}\,S(x)\times\,
\frac {1}{2}\,\tr {A^3(x)},
\ee
that is from rule $W_3$
\be
\frac {1}{8}\,S(x)\,P(x).
\ee
We thus arrive to the substitution
\be \label{substa5}
A_5(x)\to \frac{5}{6}\,(1+\frac{6}{8})\,S(x)P(x)=
\frac{35}{24}\,S(x)P(x),
\ee
which we perform in the two point correlations $A_{3,5}(x)\equiv
\av{A_3(x)\,A_5(0)}$ and $A_{5,5}(x)\equiv \av{A_5(x)\,A_5(0)}$ to get 
\bea
A_{3,5}(x) &=& \frac {35}{24}\,\av {P(x)\,P(0)\, S(0)}, \\
A_{5,5}(x) &=& \biggl (\frac {35}{24}\biggr )^2\,
    \av {P(x)\,P(0)\, S(x)\,S(0)}. 
\eea
According to Eqs. (\ref{avs}-\ref{wick3}), these averages are given by
\bea
    A_{3,5}(x)&=& \frac {35}{24}\,A_2\,A_{3,3}(x), \label{a35} \\
    A_{5,5}(x)&=& \biggl (\frac {35}{24}\biggr )^2\,
    A_{3,3}(x)\,\biggl (A_2^2+A_{2,2}(x)\biggr ). \label{a55}
\eea

As a last application, we derive the value of $A_{3,3}(0)\equiv
\av{A_3^2(x)}$. By definition
\be
A^2_3(x)=
\sum_{\alpha \,\beta \, \gamma}\,
A^\alpha A^\beta A^\gamma\, \tr \lambda^\alpha \lambda^\beta \lambda^\gamma\,
\sum_{\delta \,\epsilon \, \zeta }\,
A^\delta A^\epsilon A^\zeta\, \tr \lambda^\delta \lambda^\epsilon \lambda^\zeta,
\ee
where all the fields are taken at the same point $x$. Applying all
possible $W_2$ substitutions and using Eqs. (\ref{prod_to_trace}, 
\ref{trace_to_prod}) leads to the substitution 
\be
A^2_3(x)\to \,\frac {5S^3(x)}{64},
\ee
and averaging via Eq. (\ref{avs}) provides the final result (\ref{a33}).

\begin{figure}
\begin{center}
\includegraphics[width=12cm]{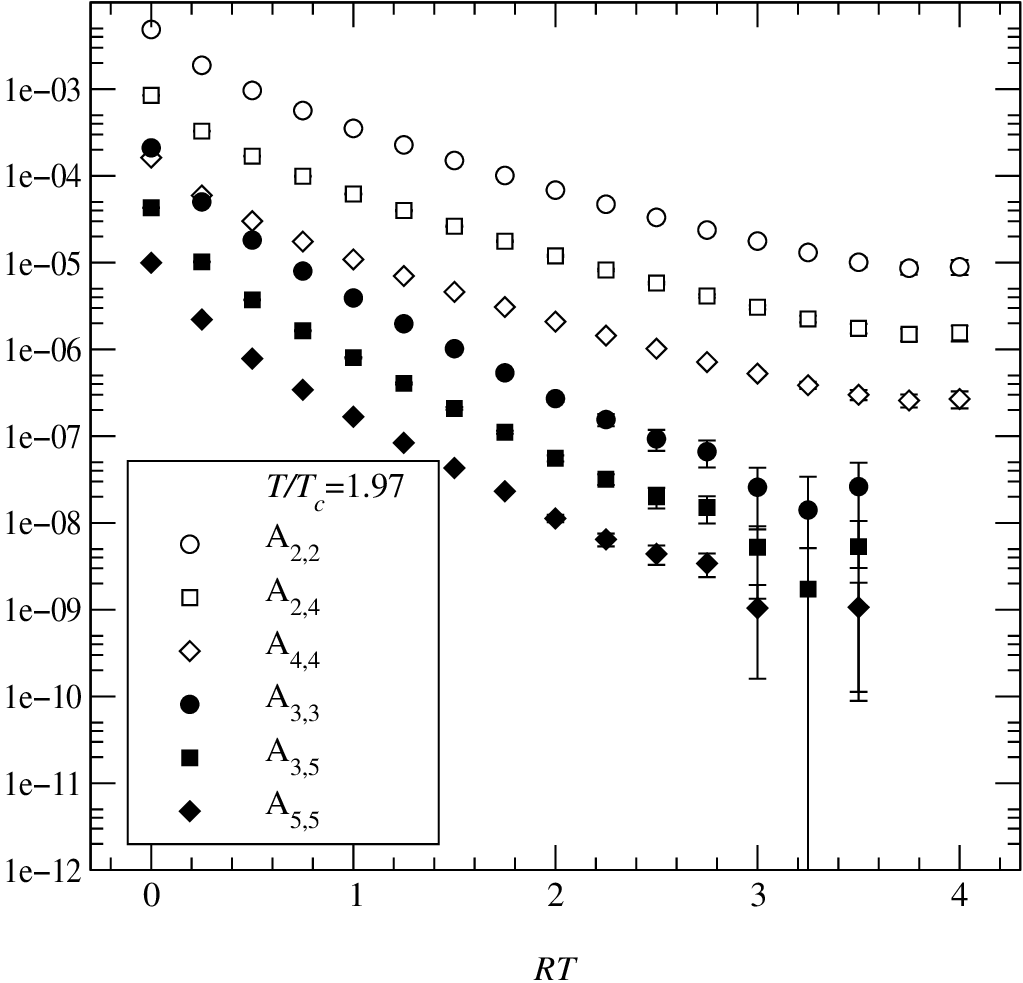}  
\end{center}
\caption[1]{The on-axis correlations $A_{n,m}(r)$ at 
$T/T_c=1.97$ ($\beta _3=29$), versus the distance in units of $1/T$. The even cases $[n,m]=[2,2],[2,4]$
and $[4,4]$ all have the same shape, and the odd cases $[3,3],[3,5]$
and $[5,5]$, again similar with each other in shape, are steeper.}
\end{figure}

\begin{figure}
\begin{center}
\includegraphics[width=12cm]{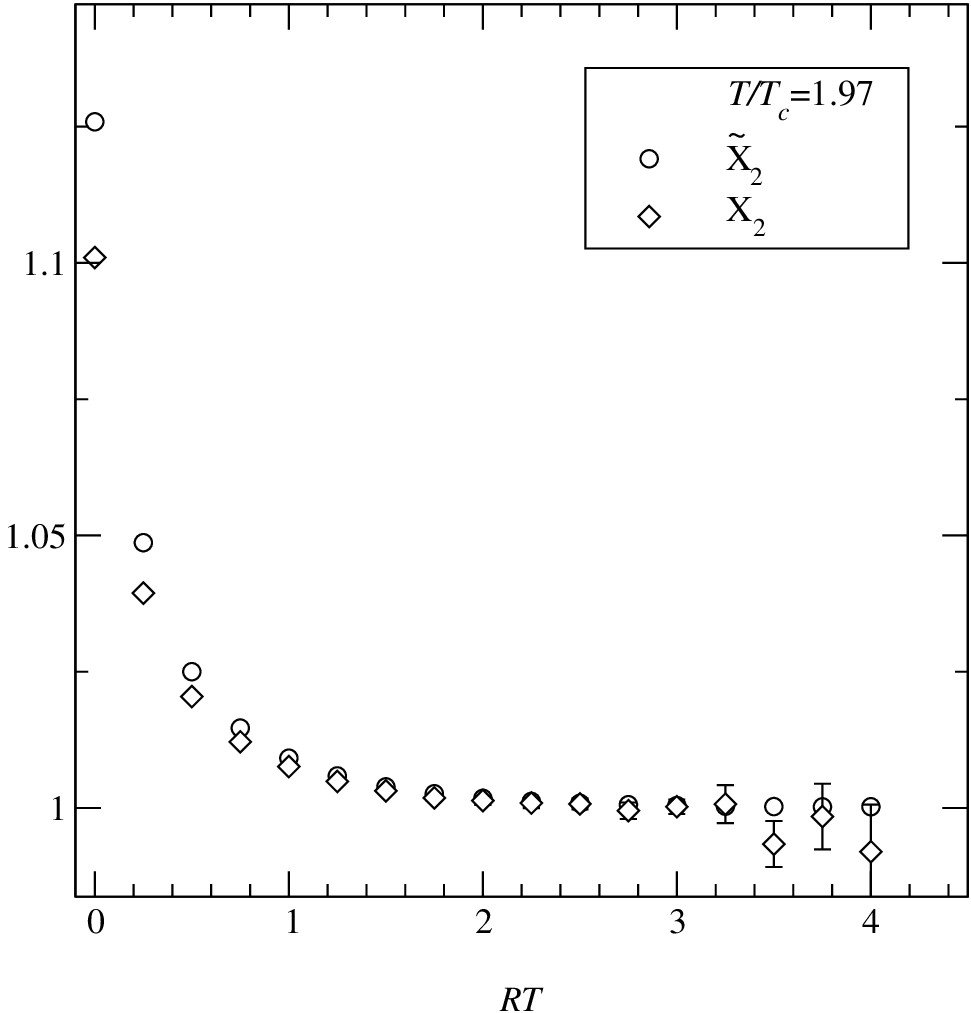}        
\end{center}
\caption[2]{Residue factorization: data for the quantity $X_2$, Eq.
(\ref{factor}) at $T/T_c=1.97$ ($\beta _3=29$) versus the distance in units of $1/T$. It approaches one at 
large distances. The quantity $\widetilde X_2$ corresponds to our interpretation 
(Section 4, Eq. (\ref{correct2})) of the deviation from one of $X_2$ at shorter 
distances.}
\end{figure}

\begin{figure}
\begin{center}
\includegraphics[width=12cm]{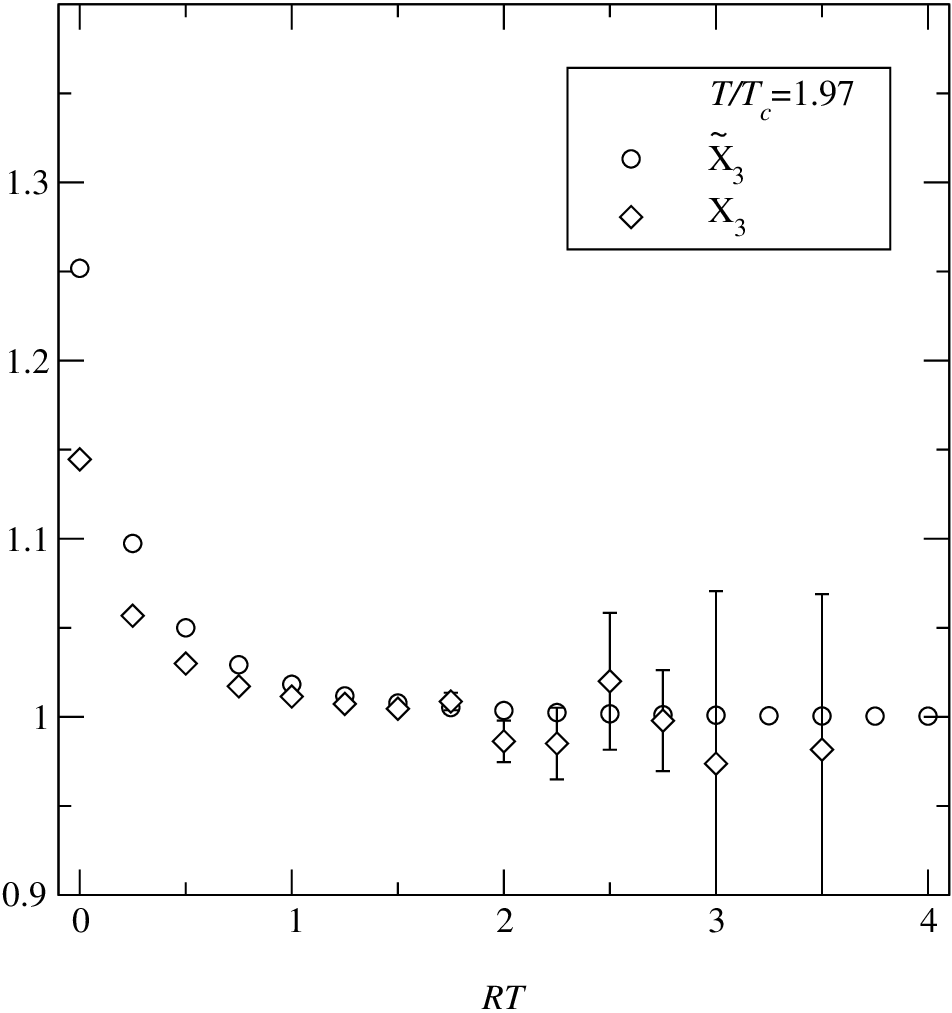}        
\end{center}
\caption[3]{Residue factorization: data for the quantity $X_3$, Eq.
(\ref{factor}) $T/T_c=1.97$ ($\beta _3=29$) versus the distance in units
of $1/T$. It approaches one at 
large distances. The quantity $\widetilde X_3$ corresponds to 
our interpretation 
(Section 4, Eq. (\ref{correct3})) of the deviation from one of $X_3$ at shorter 
distances.}
\end{figure}

\begin{figure}
\begin{center}
\includegraphics[width=12cm]{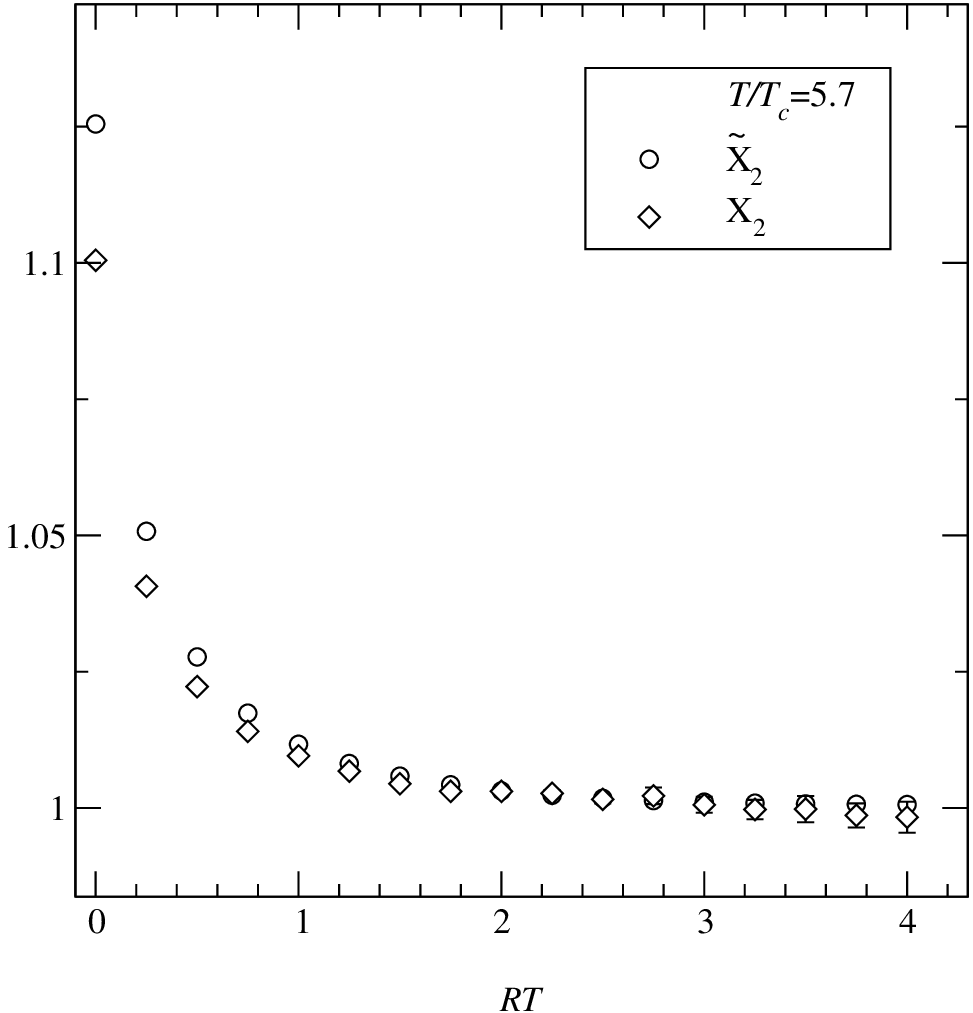}        
\end{center}
\caption[4]{Same as in Fig. 2 at $T/T_c=5.7$ ($\beta _3=84$).}
\end{figure}

\begin{figure}
\begin{center}
\includegraphics[width=12cm]{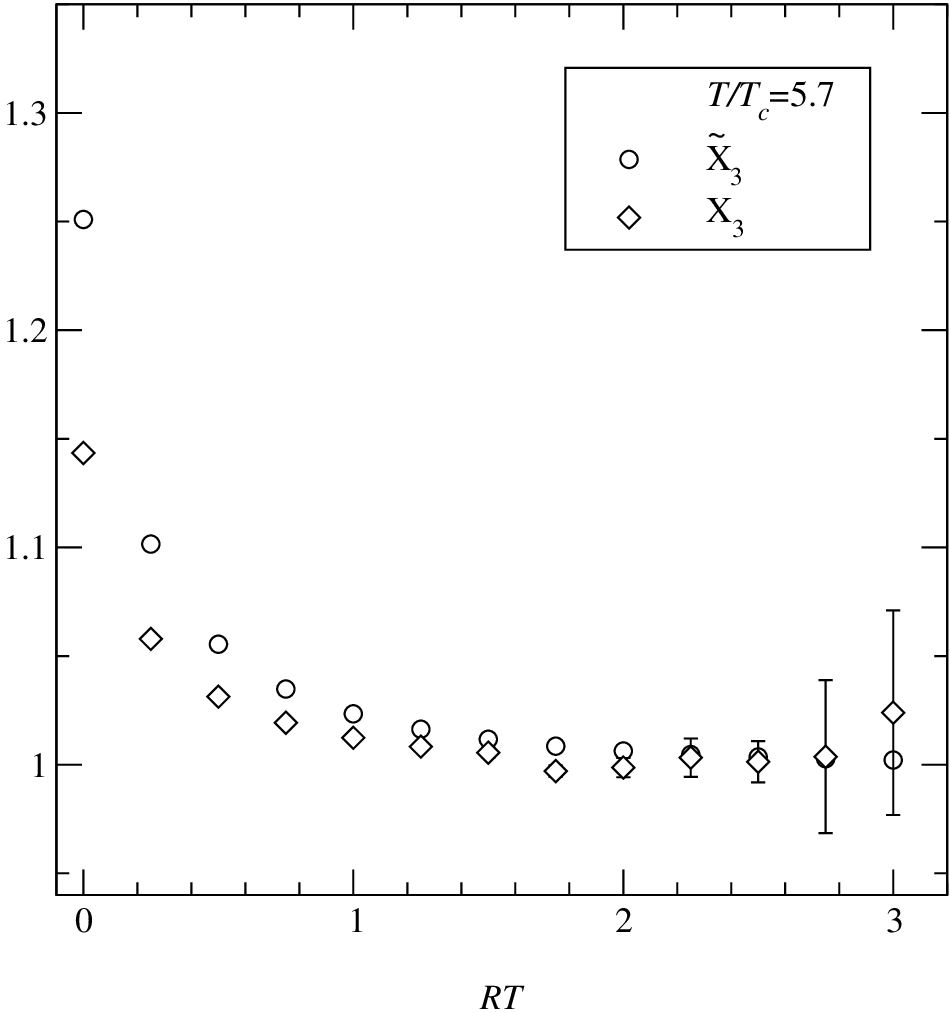}        
\end{center}
\caption[5]{Same as in Fig. 3 at $T/T_c=5.7$ ($\beta _3=84$).}
\end{figure}

\begin{figure}
\begin{center}
\includegraphics[width=12cm,angle=-90]{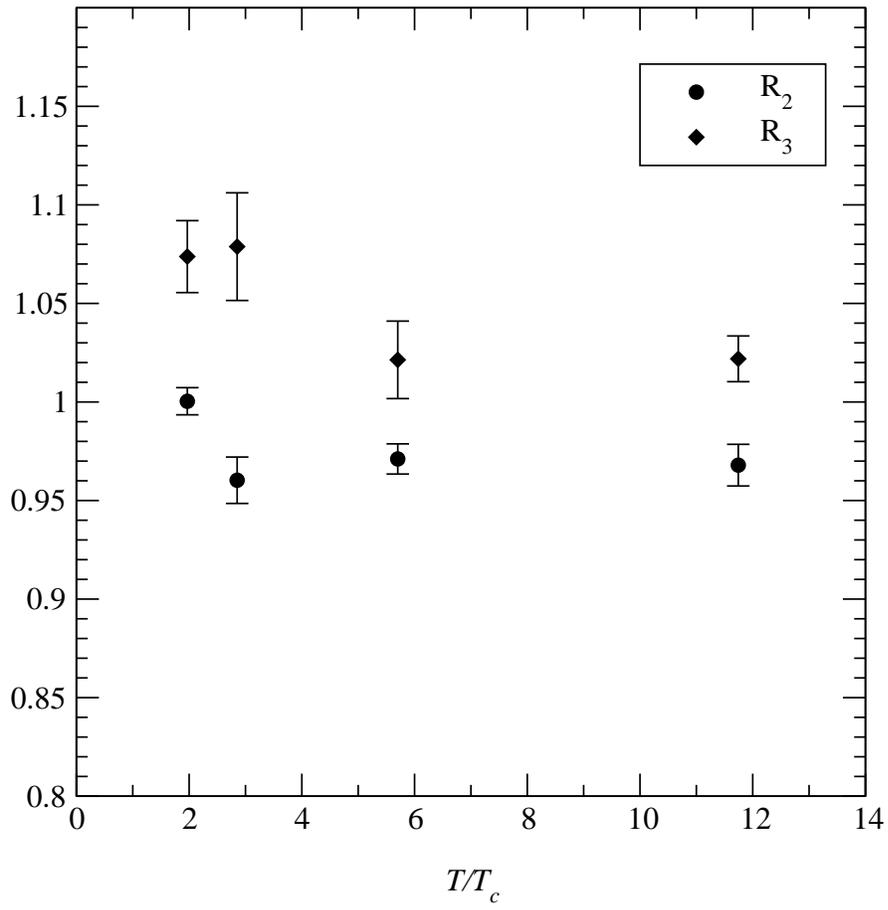}        
\end{center}
\caption[6]{The residues $R_i$ defined by Eq. (\ref{residue})
stay close to one in the whole temperature range.}
\end{figure}

\begin{figure}
\begin{center}
\includegraphics[width=12cm]{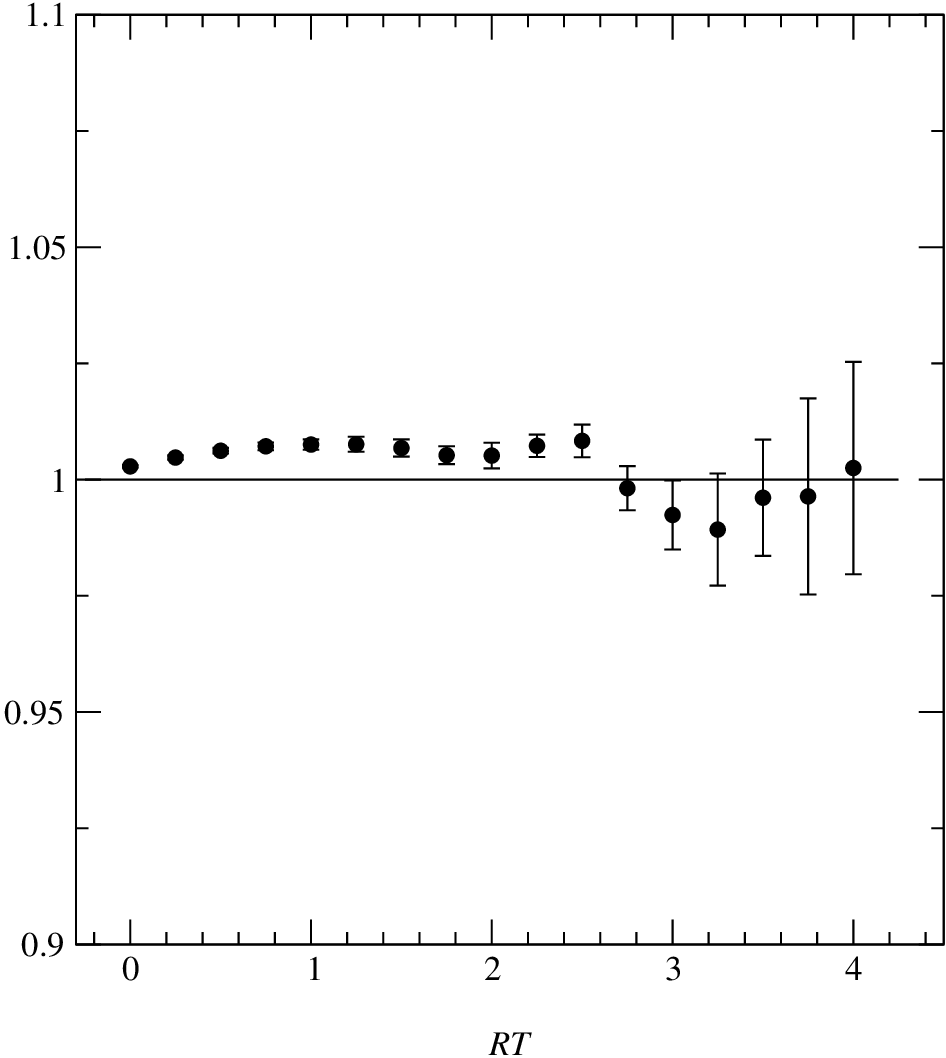}       
\end{center}
\caption[7]{ Plot of ${4A_{24}}/{5A_2A_{22}}$ at $T/T_c=1.97$ ($\beta
_3=29$). This quantity is one if Eq. (\ref{wicka24}) exactly holds.}

\end{figure}

\end{document}